\documentclass{article}
\usepackage{arxiv}
\usepackage[utf8]{inputenc} % allow utf-8 input
\usepackage[T1]{fontenc}    % use 8-bit T1 fonts
\usepackage{hyperref}       % hyperlinks
\usepackage{url}            % simple URL typesetting
\usepackage{booktabs}       % professional-quality tables
\usepackage{amsfonts}       % blackboard math symbols
\usepackage{nicefrac}       % compact symbols for 1/2, etc.
\usepackage{microtype}
\usepackage{cite}      % microtypography
\usepackage{lipsum}
\usepackage{amsmath,amssymb,graphicx}

\title{Analytical Solution for Bosonic Fields in the FRW Multiply Warped Braneworld}

\author{
  A. S. Ribeiro \thanks{I am corresponding author.}\\
  Instituto Federal do Piau\'i -- IFPI \\
  São Raimundo Nonato, Piau\'i, 64770-000, Brazil, and\\
  Departamento de F\'{\i}sica, Universidade Federal do Cear\'{a}\\
  Campus do Pici, 60455-760, Fortaleza, Cear\'{a}, Brazil.\\
   \texttt{antonius.ribeirus@ifpi.edu.br} \\
   \and 
 G. Alencar\\
  Departamento de F\'{\i}sica, Universidade Federal do Cear\'{a}\\
  Campus do Pici, 60455-760, Fortaleza, Cear\'{a}, Brazil.\\
   \texttt{geova@fisica.ufc.br} \\
   \and
 R. R. Landim\\
  Departamento de F\'{\i}sica, Universidade Federal do Cear\'{a}\\
  Campus do Pici, 60455-760, Fortaleza, Cear\'{a}, Brazil.\\
   \texttt{renan@fisica.ufc.br} 
  }

\begin{document}
\maketitle

\begin{abstract}
In this paper we find analytical solutions for the scalar and gauge fields in the Freedman-Robertson-Walker multiply warped braneworld scenario.  With this we find the precise mass spectra for these fields. We compare these spectra with that previously found in the literature for the static case.
\end{abstract}

% keywords can be removed
\keywords{Field theories in dimensions other than four \and Relativity and gravitation \and Classical field theory.}

\section{Introduction}

In the $20$'s Kaluza and Klein (KK) attempted to unify the general relativity and electromagnetism theories by adding an extra compact dimension to the ordinary spacetime. This initial effort was considered as inconsistent and was discarded by the scientific community. Extra dimension theories have attracted interest of researchers again in the early eighties of the twentieth century, with the rise of the string theory as a quantum theory of the gravitational field. New approaches of extra dimension theories have been proposed in this context, with interesting consequences in theories of elementary particles and cosmology \cite{Horava:1996ma,Horava:1995qa,Antoniadis:1998ig,Dienes:1998vh}. 

The spacetime in these higher dimensional models is generally taken as a product of a four-dimensional spacetime with $n$ compact  extra dimensions \cite{ArkaniHamed:1998rs}. While the gravity can propagates freely through the extra dimensions the particles in Standard Model (SM) are confined into spacetime in four dimensions. In add, the effective Planck scale is measured by the observers in this four-dimensional boundary as
$M^{2}_{Pl}=M^{n+2}V_{n}$, where $V_{n}$ is the volume of the compact space. If $V$ is sufficiently large, we can take TeV order to Planck scale. Thus, it is possible to remove the hierarchy between the Planck scale and $V_{n}$ weak scale. Randall and Sundrum(RS) solved this problem proposing a new scenario with a non-factorizable geometry, and five dimensions with an exponential suppression of the Planck scale \cite{Randall:1999ee,Randall:1999vf}. This naturally generates the energy of the Standard Model (SM) of particles that solves the problem of the great desert between the energy scales. The introduction of a cosmological evolution over the brane in the RS model  was considered, for example, in Refs. \cite{Flanagan:1999cu,Kim:1999ja,Binetruy:1999hy}. 

As a natural extension of the RS model, several  codimension two models were proposed  \cite{Oda:2000zj,Aghababaie:2003ar,Bonelli:2000tz,Kanti:2001vb,Giovannini:2001uf,Burdman:2005sr}. 
In this direction Debajyoti Choudhury et al proposed  multiply warped scenario  \cite{Choudhury:2006nj}.  Due to the two orbifolds, we four intersecting points, forming a picture that reminds a box. The walls of this box are $(4+1)$-branes, while in the intersecting points  there are $(3+1)$-branes. We can identify one of the intersecting points as being the Planck brane and other of these points as our (3+1) universe. The authors found the mass spectrum of the Dirac field. With this a phenomenological consequence of the this model, for flat branes,  includes an explanation of the observed hierarchy in the masses of standard model fermions\cite{Choudhury:2006nj}. The mass spectrum of the scalar and gauge  fields was found in \cite{Koley:2010za,Das:2011fb}. Whit this it is possible to  consistently describe a bulk Higgs and gauge fields with spontaneous symmetry breaking \cite{Das:2011fb}. The mass spectra of Refs. \cite{Koley:2010za,Das:2011fb} are found by considering an approximation in the equations of motion. In order to solve this Arun et al found analytical solutions for scalar and gauge fields in Refs. \cite{Arun:2015kva,Arun:2016ela}.

The generalization of the multiply warped scenario to include a FRW solution over the brane was obtained recently in Ref. \cite{Banerjee:2011wk}. However, bosonic fields has not been considered in this background. This paper is organized as follows:  in section \ref{section2} we review the static and FRW doubly warped braneworld. In section \ref{section3} we find analytical solutions to scalar and gauge fields and with this the respective mass spectra. In section \ref{section4}, using the same methodology of the previous section, we also find analytical solution for the FRW case. As it turns, in section \ref{section5} we summarize our main findings and draw some perspectives.

\section{Six-dimensional FRW Doubly Warped Spacetime}
\label{section2}

Let us review the FRW doubly warped spacetime \cite{Banerjee:2011wk}. The background is given by the manifold $M^{(1,5)}\rightarrow\left[M^{(1,3)}\times S^{1}/Z_{2}\right]\times S^{1}/Z_{2}$. Considering two consecutive orbifolds along the two extra dimensions, it is possible to obtain a configuration of a box-like spacetime \cite{Choudhury:2006nj}. 

Let us first define the notation. The spacetime coordinates are $x^{M}$, where the index are defined as $M$, $N$, with values $M,N=0,1,2,3,4,5$.  The non-compact coordinates are $x^{\mu}$,  with  $\mu,\nu=0,1,2,3$.  Finally $x^{4}=y$ and $x^{5}=z$ are the extra dimension coordinates.  The  fixed points of the orbifolds, which are the $4$-branes, are located in $x^{4}=0,\pi$ and $x^{5}=0,\pi$. In this approach two $4$-brane intersect to form the 3-brane  at the points $\left(x^{4},x^{5}\right)\rightarrow(0,0),(0,\pi),(\pi,0),(\pi,\pi)$.

To simplify the model it is relevant to assume that our universe is homogeneous and isotropic. 
With this the line element is restricted to the form 
\begin{eqnarray}
 ds^{2}&=&B^{2}\left(z\right)[A^{2}\left(y\right)\left(-dt^{2}+V^{2}(t)\delta_{ij}dx^{i}dx^{j}\right)+R^{2}dy^{2}]\nonumber\\
 &+&r^{2}dz^{2}
 \label{metric}
\end{eqnarray}
where $R$ and $r$ are the $moduli$ along the compact coordinates  $y$ and $z$, respectively. 
The total action of the background is given by
\begin{eqnarray}
S&=&S_{6}+S_{5}+S_4,\nonumber\\
S_{6}&=&\int d^{6}x\sqrt{-g_{6}}\left(\frac{M^{4}_{6}}{2}R_{6}-\Lambda_{6}\right),\nonumber\\
S_{5}&=&\int d^{4}xdydz\{\sqrt{-g_{5}}\left[\mathcal{L}_{1}-V_{1}(z)\right]\delta(y),\nonumber\\
&+&\sqrt{-g_{4}}\left[\mathcal{L}_{2}-V_{2}(z)\right]\delta(y-\pi)\}\nonumber\\
&+&\int d^{4}xdydz\{\sqrt{-\tilde{g}_{5}}\left[\mathcal{L}_{3}-V_{3}(y)\right]\delta(z)\nonumber\\
&+&\sqrt{-\tilde{g}_{5}}\left[\mathcal{L}_{4}-V_{4}(y)\right]\delta(z-\pi)\},\nonumber\\
S_{4}&=&\int d^{4}xdydz\sqrt{-g}\left[\mathcal{L}-\lambda\right].
\end{eqnarray}
The brane tensions $V_{i}$(with $i=1,...,4$), are dependent on the coordinates of the extra dimensions. The matter lagrangian of the $4$-branes are $\mathcal{L}_{i}$(with $i=1,...,4$) and $S_{4}$ is the contribution of the $3$-brane. 
The matter content of the branes is described by the perfect fluid approximation, where 
the energy density is $\rho$, and the pressure of the fluid is $p$. 

By varying the action $S$, we obtain the following Einstein's equation in six dimensions:
\begin{eqnarray}
&-&M^{4}\sqrt{-g}_{6}\left(R_{MN}-\frac{R_{6}}{2}g_{MN}\right)\nonumber\\
&=&\Lambda_{B}\sqrt{-g_{6}}g_{MN}+\nonumber\\
&-&\left[\left(T^{\gamma}_{\beta}g_{\alpha\gamma}\right)_{1}+
\left(T^{\gamma}_{\beta}g_{\alpha\gamma}\right)_{2}\right]\sqrt{-g_{4}}\delta^{\alpha}_{M}\delta^{\beta}_{N}+\label{einsteineq}\\
&-&\left[\left(\tilde{T}^{\gamma}_{\beta}\tilde{g}_{\alpha\gamma}\right)_{3}+
\left(\tilde{T}^{\gamma}_{\beta}\tilde{g}_{\alpha\gamma}\right)_{4}\right]\sqrt{-\tilde{g_{5}}}
\delta^{\alpha}_{M}\delta^{\beta}_{N}\nonumber
\end{eqnarray}
where $T$ and $\tilde{T}$ are the energy momentum tensors of the $4$-branes.

By substituting Eq. (\ref{metric}) in Eq. (\ref{einsteineq})  it is possible to obtain the following solution 
\begin{eqnarray}
V\left(t\right)&=& e^{H_{0}t} \label{eqv},\\
A\left(y\right)&=&\frac{RH_{0}}{D}\sinh\left(-D\left|y\right|+d_{0}\right)\label{eqA}, \\
B\left(z\right)&=&\frac{\cosh\left(kz\right)}{\cosh(k\pi)}\label{eqB},\\
D&=&\frac{Rk}{r\cosh(k\pi)},\label{eqD}\\
k&=&r\sqrt{-\frac{\Lambda_{6}}{10M^{4}}}\label{eqc},
\end{eqnarray}
where $H_{0}$ and $d_{0}$ are integration constants.  In the next section we compute the mass spectrum for scalar and gauge fields in this background.

\section{Analytical Solution for the Scalar Field}
\label{section3}

As said in the introduction, the mass spectrum of the scalar field for the static case was obtained approximately in Ref. \cite{Koley:2010za}. Some time latter an analytical solution was found \cite{Arun:2015kva}. In this section we consider the FRW doubly warped metric given by Eq. (\ref{metric}) and find an analytical solution for the scalar field in this background. The action of the massless scalar field is 
\begin{equation}
S=\int d^{6}x\sqrt{-g}\left(\frac{1}{2}g^{MN}\partial_{M}\Phi\partial_{N}\Phi\right),
\end{equation}
with equation of motion given by 
\begin{equation}
\frac{1}{\sqrt{-g}}\partial_{M}\left[\sqrt{-g}g^{MN}\partial_{N}\Phi\right]=0.
\label{eqn01}
\end{equation}
Making the (KK) decomposition of the scalar field through the modes
\begin{equation}
\Phi\left(x^{\mu},y,z\right)=\sum_{ij}\tilde{\phi}_{i,j}\left(x^{\mu}\right)\xi_{ij}\left(y\right)\chi_{j}\left(z\right)
\label{eqPhi}
\end{equation}
and by substituting it in Eq. (\ref{eqn01}), we find
\begin{eqnarray}
\frac{1}{V^{3}}\partial_{\mu}\left(V^{3}\tilde{g}^{\mu\nu}\partial_{\nu}\tilde{\phi}_{ij}\right)-M^{2}_{ij}\tilde{\phi}_{ij}=0,
\label{eof11}
\end{eqnarray}
\begin{equation}
 \frac{1}{r^{2}}\frac{d}{dz}\left(B^{5}\frac{d\chi_{j}}{dz}\right)=-M^{2}_{j}B^{3}\chi_{j},
 \label{eqz}
\end{equation}
and
\begin{equation}
 \frac{1}{R^{2}}\frac{d}{dy}\left(A^{4}\frac{d\xi_{ij}}{dy}\right)-M^{2}_{j}A^{4}\xi_{ij}=-M^{2}_{ij}A^{2}\xi_{ij},
 \label{eq5y}
\end{equation}
where $\tilde{g}_{00}=-1, \tilde{g}_{ij}=V^{2}(t)\delta_{ij}$ is the FRW metric and $M^{2}_{ij}$ is the effective mass seen by an observer in the 3-brane.

Let us begin by solving Eq. (\ref{eqz}). For this we first use the warp factor $B\left(z\right)$ to get
\begin{eqnarray}
 \frac{d^{2}\chi_i}{dz^{2}}+5k\tanh(kz)\frac{d\chi_{j}}{dz}+\nonumber\\ \label{eqnw}
 \left(\mbox{sech}^2(kz)r^{2}M^{2}_{j}\cosh(k\pi)\right)\chi_{j}=0.
\end{eqnarray}
In order to solve the above equation we define
\begin{equation}
\chi_j(z)=\mbox{sech}^{5/2}(kz)F_j(\omega), \quad\omega=\tanh(kz)
\end{equation}
and replace in Eq. (\ref{eqnw}) to arrive at
\begin{eqnarray}
 \left(1-\omega^{2}\right)\frac{d^{2}F_j}{d\omega^{2}}-2\omega\frac{dF_j}{d\omega}+\nonumber\\ \left[\alpha_j(\alpha_j+1)
 -\frac{25}{4(1-\omega^{2})}\right]F_j=0,
 \label{lfunc}
\end{eqnarray}
where
\begin{equation}
 \alpha_{j}=-\frac{1}{2}+\sqrt{4+\frac{r^{2}M_{j}^{2}\cosh^{2}(k\pi)}{k^{2}}}.
 \label{alfaj}
\end{equation}
The expression (\ref{lfunc}) is Legendre's associated equation with solution 
\begin{eqnarray}
\chi_{j}\left(z\right)=A_{1}P_j(z)+A_2Q_j(z),
\label{eqnchi1}
\end{eqnarray}
where
\begin{eqnarray}
P_j(z)=P^{5/2}_{\alpha_{j}}\left(\tanh(kz)\right)\mbox{sech}^{5/2}(kz)\\
Q_j(z)=Q^{5/2}_{\alpha_{j}}\left(\tanh(kz)\right)\mbox{sech}^{5/2}(kz).
\end{eqnarray}

Now we turn to solve Eq. (\ref{eq5y}). For this we replace the warp factor 
$A\left(y\right)$ in Eq. (\ref{eq5y}) to obtain
\begin{eqnarray}
\frac{d^{2}\xi_{ij}}{dy^2}-4D\coth(\theta)\frac{d\xi_{ij}}{dy}
 +\nonumber\\ \label{eqy}\left[\mbox{csch}^{2}(\theta)\frac{M^{2}_{ij}D^{2}}{H^{2}_{0}}-M_{j}^{2}R^{2}\right]\xi_{ij}=0,
\end{eqnarray}
where
\begin{equation}
 \theta=-Dy+d_{0}.
\end{equation}

Applying the  transformations
\begin{equation}
\xi_{ij}(y)=\mbox{csch}^2(\theta) G_{ij}(\tau), \quad \tau=\coth(\theta),
\end{equation}
we obtain
\begin{eqnarray} 
 \left(1-\tau^{2}\right)\frac{d^{2}G_{ij}}{d\tau^{2}}-2\tau\frac{dG_{ij}}{d\tau}+\nonumber\\ \left[\gamma_{ij}(\gamma_{ij}+1)
 -\frac{\lambda_{j}^2}{4(1-\tau^{2})}\right]G_{ij}=0,\label{Eqz}
\end{eqnarray}
with
\begin{eqnarray}
 \lambda_{j}=\sqrt{4+\frac{M^{2}_{j}R^{2}}{D^{2}}}=\alpha_j+\frac{1}{2}
 \label{betai}
\end{eqnarray}
and
\begin{equation}
 \gamma_{ij}=-\frac{1}{2}+\sqrt{\frac{9}{4}-\frac{M^{2}_{ij}}{H^{2}_{0}}}.
 \label{massive}
\end{equation}

If $M_{ij}/H_{0}>3/2$, then 
\begin{equation}
\gamma_{ij}=-\frac{1}{2}+i\sigma_{ij},
\end{equation}
where
\begin{equation} \label{shift}
\sigma_{ij}=\sqrt{\frac{M^{2}_{ij}}{H^{2}_{0}}-\frac{9}{4}}.
\end{equation}
Just as before, expression (\ref{Eqz}) is Legendre's associated equation with solution
\begin{equation}
 \xi_{ij}(y)=B_1 P_{ij}(y)+B_2 Q_{ij}(y),\label{eqnxi1}
\end{equation}
where
\begin{eqnarray}
P_{ij}(y)=Re\left[P^{\lambda_{j}}_{\gamma_{ij}}\left(\coth(-Dy+d_0)\right)\right]\mbox{csch}^2(-Dy+d_0)\\
Q_{ij}(y)=Re\left[Q^{\lambda_{j}}_{\gamma_{ij}}\left(\coth(-Dy+d_0)\right)\right]\mbox{csch}^2(-Dy+d_0).
\end{eqnarray}
 See that we have found a general solution and thus it is not necessary to take a fixed order. This occurs due to the fact that the  $\alpha_{j}$ order assumes different values for  distinct $j$, i. e., $M_{j}$ is dependent of $\alpha_{j}$. 

With our analytical solution we can obtain the precise mass spectrum of the system. For this we must apply the four boundary conditions in Eqs. (\ref{eqnchi1}) and (\ref{eqnxi1}). Observe that these conditions are applied taking the derivative of the functions that vanish at fixed points $(z=0,\pi)$, and $(y=0,\pi)$. With this we get the conditions
\begin{eqnarray}
 A_{1}P'_{j}(0)+A_{2}Q'_{j}(0)=0,\, A_{1}P'_{j}(\pi)+A_{2}Q'_{j}(\pi)=0,\\
 B_1 P'_{ij}(0)+B'_2 Q'_{ij}(0)=0,\, B_1 P_{ij}(\pi)+B_2 Q'_{ij}(\pi)=0.
 \label{As}
\end{eqnarray}
Non trivial solutions of the above equations are possible only if
\begin{eqnarray}
P'_{j}\left(0\right)Q'_{j}\left(\pi\right) - Q'_{j}\left(0\right)P'_{j}\left(\pi\right)=0,\label{Mj}\\
P'_{ij}\left(0\right)Q'_{ij}\left(\pi\right) - Q'_{ij}\left(0\right)P'_{ij}\left(\pi\right)=0.\label{Mij}
\end{eqnarray}
From Eq. (\ref{Mj}) we get the parameter $M_j$ and by using this in Eq. (\ref{Mij}) we get the mass spectrum $M_{ij}$ of the scalar field. 

To clearly illustrate the dependence of the massive modes in relation to Hubble constant we produce the table \ref{table1}. 
\begin{table}[ht]
\caption{Massive modes with Hubble constant. The $M_{pl}/H_{0}$ ratio of the scalar field  is shown in the table.
         $D=11.52$, $k=0.25$ .} 
% title of Table
\centering 
% used for centering table
\begin{tabular}{c c c c c c}
% centered columns (4 columns)
\hline\hline                        %inserts double horizontal lines
 &                   &  $M_{ij}/H_0$&                           \\   [0.5ex]
\hline\hline\\
$M_{1j}$    & $M_{2j}$      & $M_{3j}$  & $M_{4j}$    &   $M_{5j}$      &   $M_{6j}$   \\ % inserting body of the table
10.56       &    6,10       &   2,75    &    2,64     &       2,62      &       2,63     \\
22.57       &    7,65       &   7,60    &    21,21    &       22,13     &       22,86    \\
42.06       &    10,89      &   14,29   &    42,41    &       42,52     &       42,98     \\
56.06       &    21,18      &   35,16   &    63,28    &       62,36     &       63,61    \\      
77.04       &    77,04      &   77,04   &    77,17    &       77,27     &       77,49   \\
84,03       &    84,03      &   84,03   &    84,19    &       84,28     &       84,46   \\ [0.3cm]
\hline\hline                         %inserts single line
\end{tabular}
\label{table1}
% is used to refer this table in the text
\end{table}
We should point that we are considering the massless scalar field. Therefore the mass spectrum shown in the table is a correction to the one found for the static case. From Eq. (\ref{shift})
\begin{equation}\label{M94}
M_{ij}=H^{2}_{0}\sqrt{\sigma^{2}_{ij}+\frac{9}{4}}
\end{equation}
and therefore this correction is proportional to $H_0$.

\section{Analytical Solution for the Gauge Boson}
\label{section4}

In this section we must study the gauge boson. The mass spectrum of this field, for the static case, was obtained approximately in Ref. \cite{Das:2011fb}. Some time latter an analytical solution was found \cite{Arun:2015kva,Arun:2016ela}. Here we consider the FRW doubly warped metric given by Eq. (\ref{metric}) and find an analytical solution for the gauge field in this background. The action is given by
\begin{equation}
 S_{g}=\int d^{6}x\sqrt{-g}\left(-\frac{1}{4}F_{MN}F^{MN}\right),
\end{equation}
where, $g$ is the determinant of the metric given by Eq. (\ref{metric}), and  $F_{MN}=\partial_{M}X_{N}-\partial_{N}X_{M}$
is the gauge field strength. By varying the action we obtain the equation of motion 
\begin{equation}
\frac{1}{\sqrt{-g}}\partial_{M}\left[\sqrt{-g}g^{MN}g^{LK}F_{NL}\right]=0.
\label{eomg}
\end{equation}
We can choose a gauge in which $A _{4}=A_{5}=0$. Making the (KK) decomposition of the scalar field through the mode sum
\begin{equation}
X_{\rho}\left(x^{\mu},y,z\right)=\sum_{pl}\tilde{A}^{pl}_{\rho}\left(x^{\mu}\right)\eta_{pl}\left(y\right)\zeta_{l}\left(z\right),
\end{equation}
and by substituting it in Eq. (\ref{eomg}), it is possible to obtain the equations
\begin{eqnarray}
\frac{1}{V^{3}}\partial_{\mu}\left(V^{3}\tilde{g}^{\mu\nu}\partial_{\nu}\tilde{A}^{pl}_{\rho}\right)-M^{2}_{pl}\tilde{A}_{\rho}=0,\label{eof1}\\
\frac{1}{r^{2}}\frac{d}{dz}\left(B^{3}\frac{d\zeta_{l}}{dz}\right)=-M^{2}_{l}B\zeta_{l},
 \label{eqm6z}\\
\frac{1}{R^{2}}\frac{d}{dy}\left(A^{2}\frac{d\eta_{pl}}{dy}\right)-M^{2}_{l}A^{2}\eta_{pl}=-M^{2}_{pl}\eta_{pl}.
 \label{eqm5y}
\end{eqnarray}
where $M_{pl}$ is the effective mass. 

To find $M_{pl}$ we solve the Eq. (\ref{eqm6z})
 to obtain  $M_{l}$,  which must be used in Eq. (\ref{eqm5y}) to determine $M_{pl}$. Let us begin by solving Eq. (\ref{eqm6z}). First we replace the warp factor $B\left(z\right)$ to obtain
\begin{eqnarray}
 &&\frac{d^{2}\zeta_{l}}{dz^{2}}+3k\tanh(kz)\frac{d\zeta_{l}}{dz}\nonumber\\
 &+&
 [\mbox{sech}^2(kz)r^{2}M^{2}_{l}\cosh(k\pi)]\zeta_{l}=0.
 \label{eqmww}
\end{eqnarray}
Now we make the coordinate change
\begin{equation}
\zeta_{l}(y)=\mbox{sech}^{3/2}(\omega) \bar{F}_{l}(\omega),\omega=\tanh(kz).
\end{equation}
to obtain
\begin{eqnarray}
\left(1-\omega^{2}\right)\frac{d^{2}\bar{F}_{l}}{d\omega^{2}}-2\omega\frac{d\bar{F}{_l}}{d\omega}+\nonumber\\ \left[\mu_l(\mu_{l}+1)
-\frac{9}{4(1-\omega^{2})}\right]\bar{F}_{l}=0,
\end{eqnarray}
where
\begin{equation}
\mu_{l}=-\frac{1}{2}+\sqrt{1+\frac{r^{2}M_{l}^{2}\cosh^{2}(k\pi)}{k^{2}}}.
\label{alfal}
\end{equation}
The above expression is Legendre's associated equation with solution
\begin{eqnarray}
\zeta_{l}\left(z\right)=C_{1}P_{l}(z)+C_{2}Q_{l}(z),
\label{solG1}
\end{eqnarray}
where
\begin{eqnarray}
P_l(z)=P^{3/2}_{\mu_{l}}\left(\tanh(kz)\right)\mbox{sech}^{3/2}(kz)\\
Q_l(z)=Q^{3/2}_{\mu_{l}}\left(\tanh(kz)\right)\mbox{sech}^{3/2}(kz).
\end{eqnarray}

To solve Eq. (\ref{eqm5y}) we use the warp factor $A(y)$  and the transformations
\begin{equation}
\eta_{pl}(y)=\mbox{csch}(\theta) \bar{G}_{pl}(\tau),  \tau=\coth(\theta)
\end{equation}
to arrive at 
\begin{eqnarray}
\left(1-\tau^{2}\right)\frac{d^{2}\bar{G}_{pl}}{d\tau^{2}}-2\tau\frac{d\bar{G}_{pl}}{d\tau}+\nonumber\\ \left[\beta_{pl}(\beta_{pl}+1)
-\frac{\Omega^{2}_{l}}{(1-\tau^{2})}\right]\bar{G}_{pl}=0.\label{gauge}
\end{eqnarray}
In the above equation we have used the definitions
\begin{eqnarray}
\Omega_{l}=\sqrt{1+\frac{M^{2}_{l}R^{2}}{D^{2}}}=\mu_{l}+\frac{1}{2}
\label{betai1}
\end{eqnarray}
and
\begin{equation}
\beta_{pl}=-\frac{1}{2}+\sqrt{\frac{1}{4}-\frac{M^{2}_{pl}}{H^{2}_{0}}}.
\label{massive1}
\end{equation}

Besides, if $M_{pl}>H_0/2$, then
\begin{equation}
\beta_{pl}=-\frac{1}{2}+i\sigma_{pl}
\end{equation}
where
\begin{equation}
\sigma_{pl}=\sqrt{\frac{M^{2}_{pl}}{H^{2}_{0}}-\frac{1}{4}}.
\end{equation}
Again we have that Eq. (\ref{gauge}) is the Legendre's associated equation. We therefore arrive at the solution
\begin{equation}
\eta_{lp}(y)=D_1 P_{pl}(y)+D_2 Q_{pl}(y),\label{solG2}
\end{equation}
where

\begin{eqnarray}
P_{pl}(y)=Re\left[P^{\Omega_{l}}_{\beta_{pl}}\left(\coth(-Dy+d_0)\right)\right]\mbox{csch}(-Dy+d_0)\\
Q_{pl}(y)=Re\left[Q^{\Omega_{l}}_{\beta_{pl}}\left(\coth(-Dy+d_0)\right)\right]\mbox{csch}(-Dy+d_0).
\label{eqnxi}
\end{eqnarray}

From the solutions (\ref{solG1}) and (\ref{solG2}) we can obtain the mass spectrum by applying the boundary conditions at $(z=0,\pi)$ and $(y=0,\pi)$. We get 
\begin{eqnarray}
 C_{1}P'_{j}(0)+C_{2}Q'_{j}(0)=0,\, C_{1}P'_{j}(\pi)+C_{2}Q'_{j}(\pi)=0,\\
 D_1 P'_{ij}(0)+B_2 Q_{ij}(0)=0,\, D_1 P'_{ij}(\pi)+D_2 Q'_{ij}(\pi)=0.
\end{eqnarray}
Non trivial solutions of the above equations are possible only if
\begin{eqnarray}
P'_{j}\left(0\right)Q'_{j}\left(\pi\right) - Q'_{j}\left(0\right)P'_{j}\left(\pi\right)=0,\label{GMj}\\
P'_{ij}\left(0\right)Q'_{ij}\left(\pi\right) - Q'_{ij}\left(0\right)P'_{ij}\left(\pi\right)=0.\label{GMij}
\end{eqnarray}
As in the case of the scalar field, from Eq. (\ref{GMj}) we get the parameter $M_j$ and by using it in Eq. (\ref{GMij}) we get the mass spectrum $M_{ij}$ of the gauge field. 
Again we find a dependence between the effective mass of the gauge boson and the background Hubble constant given by
\begin{equation}\label{M14}
M_{pl}=H_{0}\sqrt{\sigma_{pl}^2+\frac{1}{4}}.
\end{equation}
 It is worthy to emphasize that due this all massive modes to vector field, taking cosmological effects, are lighter when compared to static case.The numerical values of massive modes are shown in table \ref{table3}.
 
\begin{table}[ht]
\caption{Massive modes with Hubble constant. The $M_{pl}/H_{0}$ ratio of the gauge boson  is shown in the table.
         $D=11.52$, $k=0.25$.} 
% title of Table
\centering 
% used for centering table
\begin{tabular}{c c c c c c}
% centered columns (4 columns)
\hline\hline                        %inserts double horizontal lines
 &                   &  $M_{pl}/H_0$&                           \\   [0.5ex]
\hline\hline
$M_{1l}$    & $M_{2l}$      & $M_{3l}$  & $M_{4l}$    &   $M_{5l}$      &   $M_{6l}$   \\ % inserting body of the table
0,895       &    0,889      &   0.881   &    0,877    &       0,876     &       0,875     \\
14,72       &    15,05      &   15,49   &    15,98    &       15,56     &       17,17   \\
28.34       &    28,52      &   28,73   &    28,99    &       29,27     &       29,60     \\
63.14       &    63,21      &   63,32   &    63,44    &       63,56     &       63,70    \\      
77.10       &    77,18      &   77,26   &    77,35    &       77,46     &       77,58   \\
84,11       &    84,16      &   84,23   &    84,31    &       84,42     &       84,53   \\ 
\hline\hline                         %inserts single line
\end{tabular}
\label{table3}
% is used to refer this table in the text
\end{table}

\section{Concluding remarks}
\label{section5}

In short, we have used a FRW doubly warped spacetime in order to investigate the scalar and gauge fields. This problem has been attached for the static case in Refs. \cite{Koley:2010za,Das:2011fb,Arun:2015kva,Arun:2016ela}. For a FRW background (see Ref \cite{Banerjee:2011wk}), the study of bosonic fields is lacking. 

For the scalar and gauge fields we find that the equations of motion can be transformed to a Legendre's associated equation. Therefore analytical solutions can be found and are given in Eqs. (\ref{eqnchi1}), (\ref{eqnxi1}),(\ref{solG1}) and (\ref{solG2}). With these solutions and by imposing the boundary conditions we obtain the mass spectrum of both fields. Some of the mass modes are given in tables \ref{table1} and \ref{table3}. An interesting result is that in both cases the massive modes(Eqs. (\ref{M94}) and (\ref{M14})),  are given by  
\begin{equation}
M_{pl}=H_{0}\sqrt{\sigma_{pl}^2+\frac{n}{4}},
\end{equation}
where $n=9$ or $n=1$ for the scalar and gauge fields respectively. Therefore the mass tower, in the cosmological scenario of Ref. \cite{Banerjee:2011wk}, are proportional to the parameter $H_0$. 

We should point that the above protocol can be used to investigate other bosonic fields in the FRW multiply warped background. For example, the graviton mass spectrum has been found for the static case in Ref. \cite{Arun:2014dga}. Another possibility is the $p$-form field, which has been studied  in Refs. \cite{Alencar:2010hs,Alencar:2010vk,Alencar:2018cbk} in other backgrounds. In the near future we must present analytical solutions of these fields in FRW multiply warped background.

\section{Acknowledgments}

The authors would like to thanks Alexandra Elbakyan and sci-hub, for removing all barriers in the way of science. We acknowledge the financial support provided by the Conselho Nacional de Desenvolvimento Cient\'\i fico e Tecnol\'ogico (CNPq) and Funda\c c\~ao Cearense de Apoio ao Desenvolvimento Cient\'\i fico e Tecnol\'ogico (FUNCAP) through PRONEM PNE0112- 00085.01.00/16

\bibliographystyle{unsrt}  

%\section*{References}

\bibliographystyle{elsarticle-num}
\bibliography{references}

\end{document}